\documentstyle[12pt,aasms4,psfig]{article}
\def\etal{et al. }

\begin{document}

\title
{\bf Dependence of the Fundamental Plane Scatter on Galaxy Age}

\author{Duncan A. Forbes, Trevor J. Ponman, Richard J. N. Brown}
\affil{School of Physics and Astronomy, University of Birmingham,
Birmingham, B15 2TT}
\affil{Email: forbes, tjp, rjnb@star.sr.bham.ac.uk}

\begin{abstract}

The fundamental plane (FP) has an intrinsic
scatter that can not be explained purely by observational errors. 
Using recently available age estimates for nearby early type galaxies, we
show that a galaxy's position relative to the FP depends on
its age. In particular, the mean FP corresponds to ellipticals with an age
of $\sim$10 Gyr. Younger galaxies are systematically brighter with higher
surface brightness relative to the mean relation. Old ellipticals form an
`upper envelope' to the FP. For our sample of mostly non--cluster galaxies,
age can account for almost half of the scatter in the B band FP. Distance
determinations based on the FP may have a systematic bias, if the mean age
of the sample varies with redshift.

We also show that fundamental plane residuals, B--V colors and Mg$_2$ line
strength are consistent with an ageing central burst superposed on an
old stellar population. This reinforces the view that these 
age estimates are tracing the last major episode of star formation induced by
a gaseous merger event. We briefly 
discuss the empirical `evolutionary tracks' of
merger--remnants and young ellipticals in terms of their key observational
parameters. 

\end{abstract}


\section{Introduction}

The `merger hypothesis' (Toomre \& Toomre 1972) suggests
that an elliptical galaxy will eventually result from the merger of two
spiral galaxies. A more generalised paradigm for elliptical galaxy
formation is that of hierarchical clustering and merging (HCM). Under this
framework both stellar and gaseous mergers may form an elliptical
(e.g. Bender \etal 1992). 
Other factors such as the progenitor structure, mass ratios,
orbital parameters, subsequent star formation and feedback processes will
also contribute to the nature of the final remnant. Given this range of
possible initial conditions it is remarkable that the
eventual products -- elliptical galaxies -- are so homogeneous, with their
properties scaling in a well defined manner. 

The primary scaling relation for elliptical galaxies 
involves three parameters which relate 
galaxy structure (e.g. effective radius r$_e$ and  
surface brightness within the effective radius $\mu_e$) to dynamics
(e.g. central velocity dispersion $\sigma_o$). This 3--parameter relation
is called the fundamental plane (Djorgovski \& Davis 1987; Dressler \etal
1987). Ellipticals do not fill this plane entirely but
rather a band  within it (Guzm\'{a}n \etal 1993). 

The existence of the fundamental plane (FP) 
and its small scatter ($\sim$ 0.1 dex) provide tight constraints on
elliptical galaxy formation and evolution (Bender \etal 1992; 
Guzm\'{a}n \etal 1993; Renzini \& Ciotti 1993; Zepf \& Silk 1996; Pahre
\etal 1995, 1998). 
Although small, the FP scatter is about
twice that expected from measurement errors alone (
J$\o$rgensen \etal 1996). Application of the FP for  
distance determination has provided a strong impetus to understand
the scatter about the FP and its various projections (e.g. Guzm\'{a}n \&
Lucey 1993). A number of studies (e.g. Gregg 1992; 
J$\o$rgensen \etal 1993; Prugniel \& Simien 1996)
indicate that variations in stellar populations 
are responsible for
some of the scatter. By including the additional scaling relations
involving color and metallicity, 
Prugniel \& Simien (1996) were able to
confirm that blue, low metallicity ellipticals deviate systematically from the
FP. Such galaxies probably possess younger stellar populations than the
norm. 

Ellipticals with blue colors and 
relatively low Mg$_2$ line 
indices are structurally disturbed galaxies (Schweizer
\etal 1990). Furthermore the correlation of color and line strength with
a disturbance parameter has been interpreted as due to star
formation associated with a gaseous merger event (Schweizer \& Seitzer
1992). Thus evidence is accumulating
that a galaxy's position relative to the FP is related to the time since
its last major burst of star formation, which in turn is due to a gas--rich
encounter with another galaxy. 
Showing that the FP scatter varies  
systematically with post--merger age would help confirm this picture and
provide a general consistency test of the merger hypothesis.

In this {\it Letter} we compare the residuals from the FP and other scaling
relations with recent age
estimates for 88 early type galaxies and find a strong correlation. 
We interpret the FP residual -- age trend in terms of an ageing burst of
star formation and make 
an initial empirical attempt at describing the 
evolution of elliptical galaxies in terms of their key observational
parameters. A future paper (Forbes \etal 1998) will further explore the 
evolution of merger--remnants and the implications for elliptical galaxy
formation. 

\section{The Datasets}

The main dataset used in this study is the 
compilation of over 400 (mostly non--cluster) 
early type galaxies from Prugniel \& Simien (1996).
They defined the residual from the elliptical galaxy FP as 
R($\sigma_o, M_B, \mu_e) = 2log(\sigma_o) + 0.286M_B + 0.2\mu_e -
3.101$. We also use their  
distance modulus (for H$_{\circ}$ = 
75 km s$^{-1}$ Mpc$^{-1}$); total blue magnitude (B$_T^o$); 
average B--V color
index; magnesium line strength index (Mg$_2$) and 
central velocity dispersion ($\sigma_o$). 
The typical galaxy in this sample has M$_B$ $\sim$ --20.5 and lies at a
distance of $\sim$30 Mpc. 

Estimating the age of elliptical galaxies from their stars 
has proved difficult due to the
age--metallicity degeneracy of old stellar populations. Recently however
this degeneracy has been broken by the combination of stellar spectroscopy
and new models from the Santa Cruz group 
(e.g Gonzalez 1993; Worthey 1994). 
Using spectroscopic ages from the Santa Cruz group (Trager \etal 1998) and
other workers (i.e. Kuntschner \& Davies (1998), Tantalo \etal (1998) and 
Mehlert \etal 1997) we have compiled a list of galaxy ages. 
Random errors on the age estimates are $\sim$ 20\%. 
An alternative method for estimating galaxy age that does not use
spectroscopy is that of Schweizer \& Seitzer (1992). They 
associated morphological disturbance, via a star formation
model, to a merger age. When a spectroscopic age wasn't available, 
we included morphological ages. 
We note that strictly
speaking the spectroscopic ages are central luminosity--weighted ages
presumably representing the last episode of star formation, 
whereas the `disturbance' ages are claimed to represent the time since the
last merger event. 

Our sample consists of 65 ellipticals and 24 S0s with age estimates; all
but two are also in the 
Prugniel \& Simien (1996) list.
These galaxies are mostly 
in non--cluster environments and have a similar average luminosity and
distance to the Prugniel \& Simien sample. 

\section{Deviations from the Fundamental Plane with Galaxy Age}

In Figure 1, we show the FP residuals calculated by Prugniel \& Simien (1996)
plotted against galaxy metallicity and age. For the former, we show the
galaxy metallicity (Z/Z$_{\odot}$) derived by Trager \etal (1998) 
for their data, which consitutes 70\% of our 
spectroscopic ages.
Figure 1a shows that the scatter about the FP is not strongly
dependent on metallicity for this sample of mostly field
ellipticals. 
On the other hand, Figure 1b shows that {\it the scatter about 
the FP is correlated with age. } This is true for both the Trager \etal
subsample and for all 89 galaxies. 
Prugniel \& Simien have shown that FP residuals are 
probably due to stellar population differences, namely ellipticals with
negative residuals (i.e. young ellipticals) tend to be
brighter with a higher surface brightness. 
Thus we can provide strong support for the idea that 
an elliptical's position relative to the FP is 
dependent on the time since its last starburst. 
The mean residual value for our sample 
is R($\sigma_o, M_B, \mu_e)$ = --0.07 $\pm$ 0.02. 
This indicates that it is on average, slightly younger than the full 
Prugniel \& Simien sample. 

We have included four galaxies (NGC 2865, 3156, 3921, 7252) in Figure 1b  
which are good candidates for being
the remnant of a gaseous merger about 1 Gyr ago. 
The location of these late stage
mergers close to young ellipticals in this figure 
suggests that this burst was induced by a gaseous merger event, which may
have accompanied the formation of the elliptical itself.

In order to estimate how much of the FP scatter is due to age effects
we have calculated a least--squares fit to the E and S0 
galaxies in Figure 1b. 
The variance about this fit is 0.025, whereas the variance
assuming no FP dependence on galaxy age would be 0.038. 
This indicates that at least 35\% (it would be more if observational errors
weren't present) of the FP variance in the B band is due to age
effects. If we exclude S0s, then 43\% of the variance is due to age.  
Thus the introduction of age as a fourth parameter 
can be used to better characterise any individual galaxy. 
This result has implications for using the FP for distance determination. 
In particular, any sample for which
the mean age varies with galaxy redshift will be systematically biased. 
Comparisons between field and cluster FPs may also be affected by mean age
differences. 
Prugniel \& Simien (1996) found that stellar population, rotation support
and mass profile effects can fully explain the tilt of the FP. In paper II
we show that the FP tilt includes a weak age dependence. This age dependence
contributes to the stellar population component found by Prugniel \& Simien.

From Figure 1b it appears that as young ellipticals evolve they approach the
FP. In principle, this could be due to a combination of reduced $\mu_e$,
increased $\sigma_o$ or global fading. 
In order to understand which of these evolutionary changes are 
responsible we require additional information. 
In the next section, we incorporate a galaxy's
B--V color and Mg$_2$ line strength to trace the evolution of a
merger--induced starburst. 
Interestingly, ellipticals have zero residuals at about 10 Gyr and continue
towards positive residuals as they get older. This suggests that we should
consider the FP as having an envelope defined by the oldest ellipticals,
and that the current FP relation represents typical $\sim$ 10 Gyr
ellipticals.

\section{Central Starburst Evolution}

We now examine the evolution of merger--remnants and young
ellipticals in more detail by using the correlations of 
B--V and Mg$_2$ with galaxy mass. For elliptical galaxies (with M$_B <
-18$ and $\sigma_o > 70$ km s$^{-1}$) 
from the full Prugniel \& Simien (1996) sample, we have determined
the best linear fit to the 
scaling relations of B--V and Mg$_2$ with M$_B$ and log $\sigma_o$.
The residuals about these correlations take the form 
of $R(Y,X) = Y - a~X - b$. 
The coefficients from all four scaling relations are given in
Table 1. 
 
The log residuals from these scaling relations, for our 88 sample 
galaxies are shown in Figure 2, again plotted against log galaxy age. 
Young ellipticals tend to have negative residuals in the sense that
they are bluer and have weaker line strengths for a given mass.
Figure 2 also shows that 
late stage mergers fit smoothly into the chronological trend. 

Are the luminosity, color and line strength 
trends consistent with the fading of a
merger--induced starburst ? To answer this question in detail 
would require a sophisticated model which includes not only properties of
the starburst but also dynamical changes for a 
range of initial conditions. When created, such models
could then be used to predict the evolution of $\mu_e$ and
$\sigma_o$ in a merger--remnant. However here we are interested in
understanding the broad trends which are perhaps best illustrated by using
a simple model with reasonable assumptions. 
We have chosen 
an instantaneous burst model from Bruzual \& Charlot (1993), 
as the merger--induced starburst occurs over a very short
timescale (i.e. few Myr) compared to the merger timescale. 
The model has a 
Salpeter IMF with solar metallicity. 
We have assumed that the
burst dominates the central region  
but contributes only 10\% of the total
galaxy mass (e.g. Hibbard \& van Gorkom 1996). 
The remaining 90\% of the galaxy is taken to be an 
old stellar population
formed at high redshift, so that it 
has the same age for each galaxy in Figure 2. 
Its properties are chosen to be  appropriate for a typical old
elliptical. 
Unlike the global quantities B--V and M$_B$, Mg$_2$ is typically measured
in the central region only and we assume a 100\% contribution from the burst.
We have held $\sigma_o$ constant as 
merger simulations (e.g. Barnes 1992; Navarro 1990) indicate that there is
little or no internal dynamical evolution after the merger is complete. 

To summarise, the evolution of the burst 
magnitude, color and Mg$_2$ index are described by the Bruzual \& Charlot
model, while the old galaxy stars have a constant 
total magnitude, color and Mg$_2$ index.  The central 
velocity dispersion is fixed. 
The final model has the
following contributions: M$_B$ 10\% burst by mass 
90\% fixed, B--V 10\% burst 90\%
fixed and Mg$_2$ 100\% burst. 
These models can then be used to crudely track the evolution of an elliptical
following a central 
starburst in the parameter space of B--V, Mg$_2$, M$_B$ and
$\sigma_o$. Thus we can form evolutionary tracks 
(shown by solid lines) in the residual plots of
Figure 2. 
From Figure 1b we found that a typical ellipticals have zero FP residuals 
at $\sim$ 10
Gyr. We have therefore chosen to normalise the models 
to zero residual at 10
Gyr with respect to each of the scaling relations, thus defining 
the properties of the old stellar population.

Figure 2 shows that age effects the scatter in all four scaling relations
presented. The data are consistent with the 
steady reddening, increasing line strength and global fading 
of an ageing burst of star formation. 
Thus it appears that 
the color, luminosity and line strength trends can be understood purely
in terms of evolving stellar populations.

We have implicitly assumed in our models 
that the burst fraction and metallicity are both
independent of post--merger age. At early epochs it is quite plausible that
the progenitor galaxies would be more gas--rich than today. We would also
expect this gas to be relatively metal--poor. These differences may however
have little effect on the model track as they tend to work in opposite
directions. Increasing the burst strength steepens the model curve, while a
lower metallicity for the burst population tends to flatten it. 

\section{Concluding Remarks}

In this {\it Letter} we have shown that the scatter about the fundamental
plane (FP), at least for non--cluster galaxies, is largely due to variations in
galaxy age (which probably reflects the time since the last major burst of
star formation). This supports the view that field ellipticals span a wide
range in age with only small differences in metallicity (Trager
\etal 1998). On the other hand, galaxies in the Fornax cluster appear to be
roughly coeval but with a range of metallicities (Kuntschner \& Davies
1998). We speculate that the FP scatter for cluster ellipticals may reveal
a stronger metallicity rather than age dependence. 
Starburst models with a wider range of 
parameters (e.g. different metallicities,
non--Salpeter IMF etc) may provide additional insight into the evolution of
merger--remnants with respect to galaxy scaling relations. 

\noindent{\bf Acknowledgments}\\
We thank Mike Merrifield, Stephen Helsdon and Phil James 
for helpful discussions. We also thank the referee for suggestions that
have improved the paper.\\

\noindent{\bf References}\\
Barnes, J. E. 1992, ApJ, 393, 484\\
Bender, R., Burstein, D., \& Faber, S. M. 1992, ApJ, 399, 462\\
Bruzual, G. A., \& Charlot, S. 1993, ApJ, 405, 538\\
Djorgovski, S., \& Davis, M. 1987, ApJ, 313, 59\\
Dressler, A., Lynden-Bell, D., Burstein, D., Davies, R. J., Faber, S. M., 
Terlevich, R. J., \& Wegner, G. 1987, ApJ, 313, 42\\
Forbes, D. A., \etal 1998, in preparation\\
Gonzalez, J. J. 1993, Ph.D Thesis, University of California, Santa Cruz\\
Gregg, M. D. 1992, ApJ, 384, 43\\
Guzman, R., \& Lucey, J. R. 1993, MNRAS, 263, L47\\
Guzman, R., Lucey, J. R., \& Bower, R. G. 1993, MNRAS, 265, 731\\
Hernquist, L., Spergel, D., \& Heyl, J. 1993, ApJ, 416, 415\\
Hibbard, J. E., \& van Gorkom, J. 1996, AJ, 655, 111\\
Kuntschner, H., \& Davies, R. L. 1998, MNRAS, 295, 29\\
J$\o$rgensen, I., Franx, M., \& Kjaergaard, P. 1993, ApJ, 411, 34\\
J$\o$rgensen, I., Franx, M., \& Kjaergaard, P. 1996, MNRAS, 280, 167\\
Mehlert, D., Bender, R., Saglia, R. P., \& Wegner, G. 1997, astro-ph/9709295\\
Navarro, J. 1990,  MNRAS, 242, 311\\
Navarro, J., Frenk, C., \& White, S. D. M. 1997, ApJ, 490, 493\\
Pahre, M. A., Djorgovski, S. G., \& de Carvalho, R. R. 1995, ApJ, 453,
L17\\
Pahre, M. A., Djorgovski, S. G., \& de Carvalho, R. R. 1998, preprint\\
Prugniel, P., \& Simien, F. 1996, A\&A, 309, 749\\
Renzini, A., \& Ciotti, L. 1993, ApJ, 416, L49\\
Schweizer, F., Seitzer, P., Faber, S. M., Burstein, D., Dalle Ore, C. M., \& 
Gonzalez J. J. 1990, ApJ, 364, L33\\
Schweizer, F., \& Seitzer, P. 1992, 104, 1039\\
Tantalo, R., Chiosi, C., \& Bressan, A. 1998, A\&A, 333, 419\\
Toomre, A., \& Toomre, J. 1972, ApJ, 178, 623\\
Trager, S. C., Faber, S. M., Gonzalez, J. J., \& Worthey, G. 1998, in
preparation\\ 
Weil, M., \& Hernquist, L. 1996, ApJ, 460, 121\\
Worthey, G. 1994, ApJS, 95, 107\\
Zepf, S., \& Silk. J. 1996, ApJ, 466, 114\\

\newpage
\noindent
Fig. 1 -- 
Residuals from the fundamental plane of elliptical galaxies versus galaxy
metallicity and age. 
Ellipticals are represented by filled circles, S0s by open circles
with a horizontal line and late stage mergers by a star symbol. 
1a) shows the metallicity for the subsample of spectroscopic ages
from Trager \etal (1998). Two crosses indicate the predicted 
fundamental plane residual for Z = 1 Z$_{\odot}$ and 2.5 Z$_{\odot}$
metallicity. There is no strong dependence of residual on galaxy
metallicity. 1b) shows a correlation between fundamental plane 
residual and galaxy age. 
The dotted line represents the best fit to the elliptical and S0 galaxies. 
Young galaxies have negative residuals 
but evolve to have zero residual at 
$\sim$ 10 Gyr. Older galaxies have positive residuals. 
At least a third of the variance  
within the B band fundamental plane can
be explained as an age effect. \\

\noindent
Fig. 2 -- 
Residual from various elliptical galaxy scaling relations 
versus galaxy age. The same symbols are used as in Figure 1. 
The solid line shows the evolutionary track for a galaxy 
which has undergone an instantaneous, solar
metallicity starburst involving 10\% of the total galaxy mass (the
remaining 90\% is assumed to be a non--evolving old stellar population).  
The velocity dispersion is held constant. The tracks have 
been normalised to zero residual at 10 Gyr. See text for model details.
The evolution from late stage mergers to old ellipticals is 
generally consistent with expected changes in stellar population 
following a merger--induced starburst.

\newpage

\begin{figure*}[p]
\centerline{\psfig{figure=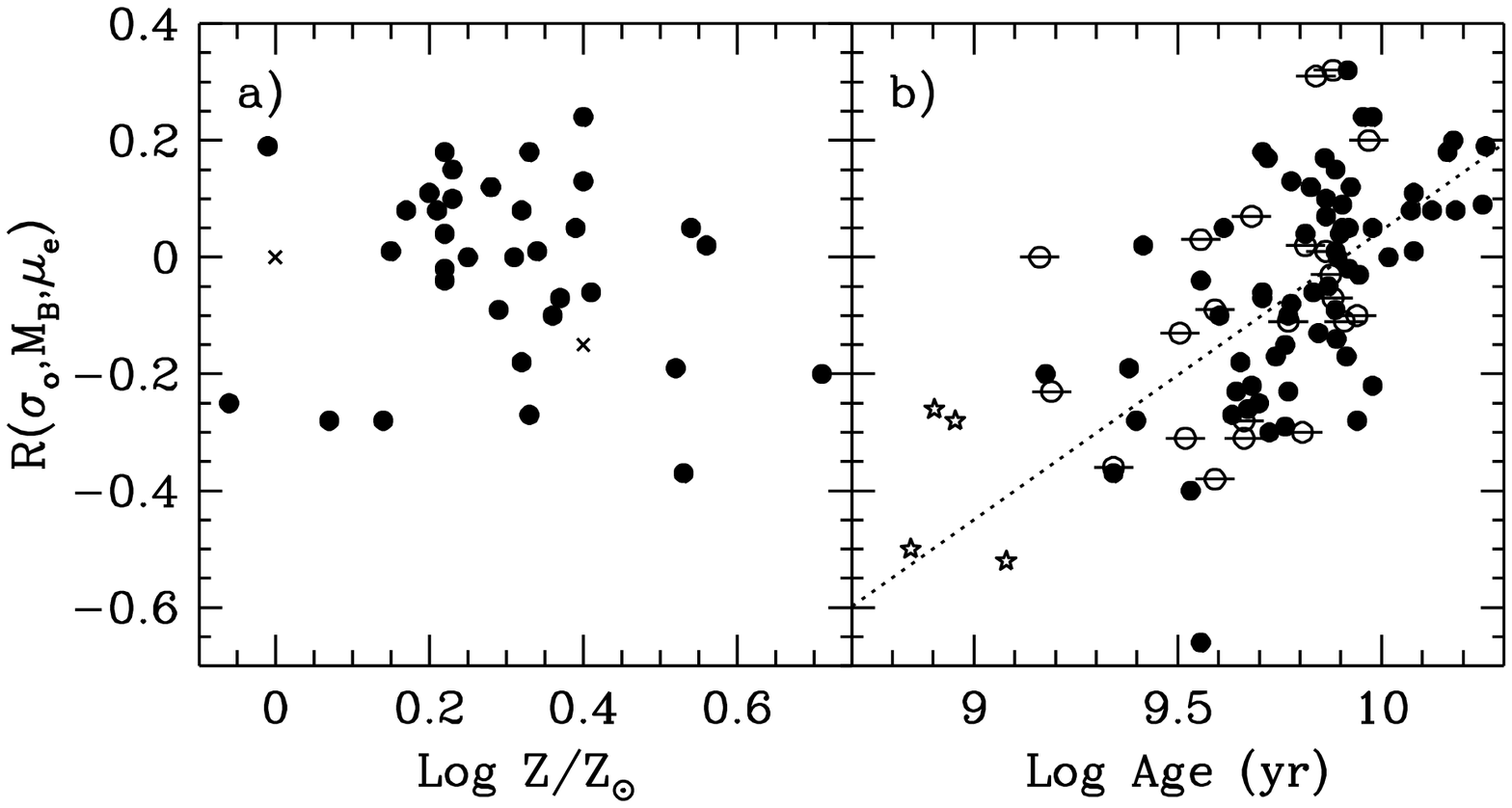}}
\caption{\label{fig1}}
\end{figure*}

\newpage

\begin{figure*}[p]
\centerline{\psfig{figure=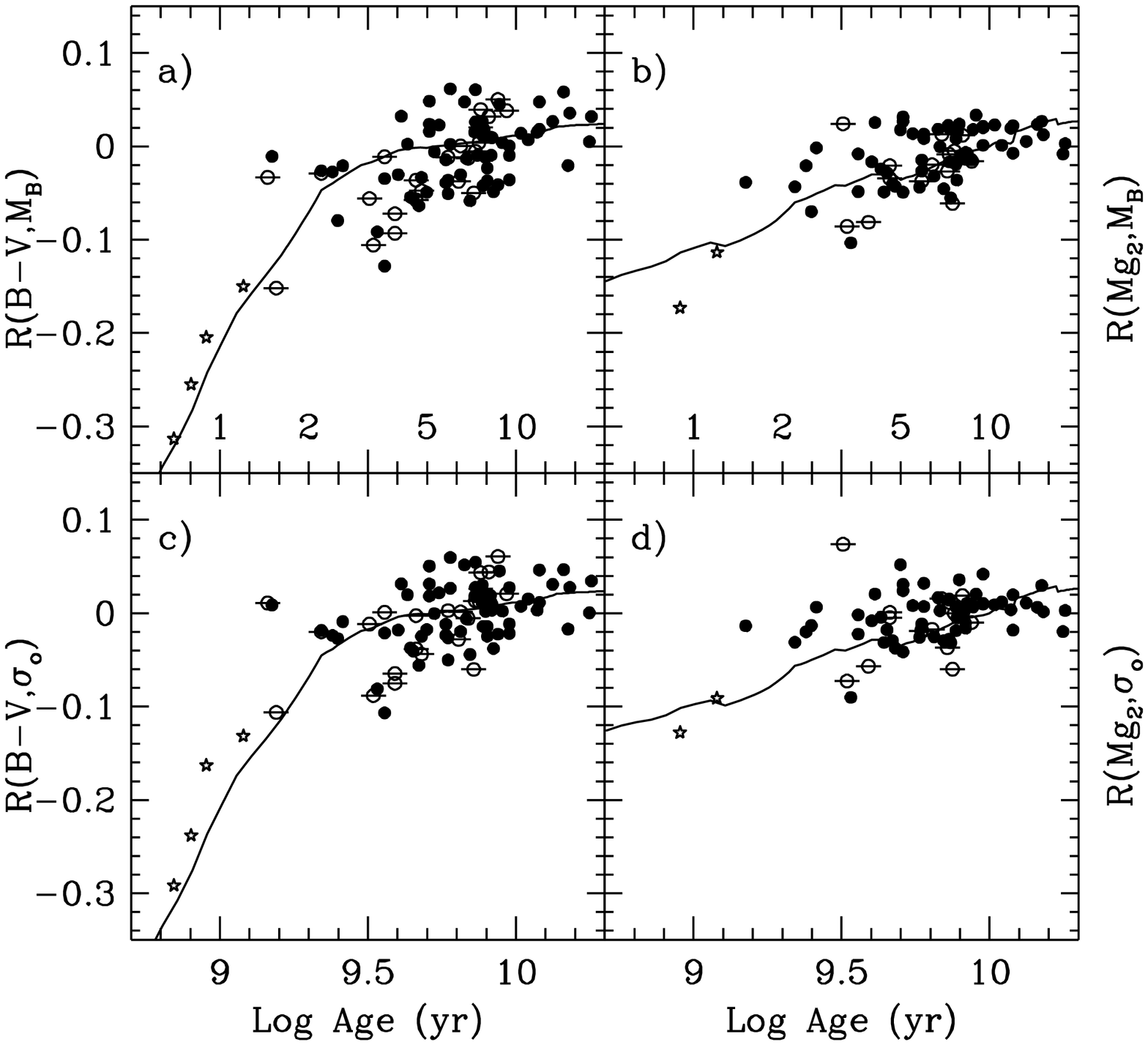}}
\caption{\label{fig2}}
\end{figure*}

\end{document}